\documentstyle[12pt, epsfig]{article}

\title{On the solitons of the Chern-Simons-Higgs model. }
\author{Wifredo Garc\'{\i}a Fuertes \\Departamento de F\'\i sica, Facultad de Ciencias, \\ Universidad de Oviedo \\ SPAIN\\and\\Juan Mateos Guilarte \\Departamento de F\'\i sica, Facultad de Ciencias, \\ Universidad de Salamanca \\ SPAIN}
\date{}
\newcommand{\beq}{\begin{equation}}
\newcommand{\eeq}{\end{equation}}
\newcommand{\bdm}{\begin{displaymath}}
\newcommand{\edm}{\end{displaymath}}
\begin{document}
\maketitle
\begin{abstract}
Several issues concerning the self-dual solutions of the Chern-Simons-Higgs model are addressed. The topology of the configuration space of the model is analysed when the space manifold is either the plane or an infinite cylinder. We study the local structure of the moduli space of self-dual solitons in the second case by means of an index computation. It is shown how to manage the non-integer contribution to the heat-kernel supertrace due to the non-compactness of the base space. A physical picture of the local coordinates parametrizing the non-topological soliton moduli space arises .
\end{abstract}
\pagebreak
\section{Introduction. }
For several years, the self-dual structure of the Chern-Simons-Higgs model \cite{bib: jw, bib: hky} has attracted the attention of theorists, giving rise to a large body of work on the subject both from the physical and mathematical points of view (see \cite{bib: dunne} for a review and a comprehensive list of references). The CSH model belongs to a family of generalized Abelian Higgs models \cite{bib: ln} and one of the most notorious novelties with respect to the old AHM is the simultaneous appareance of topological and non-topological defects in the theory. While the former have been thoroughly investigated \cite{bib: wg}, the latter are more difficult to deal with and their nature remains less clear. In this paper we elaborate on three different aspects of the issue that remain to be fully developed:

1. The topological structure of the configuration space of the CSH model is richer than that arising from the AHM. We shall address this question carefully both on the plane and on the cylinder in a way slightly different from the mainly analytical classification given in \cite{bib: jlw}. As a bonus, we find analytical self-dual solutions on the cylinder in a physical context, in contrast with the Painlev\'e analysis of reference \cite{bib: schi}. These solutions are the finite energy counterparts of the domain walls of Jackiw, Lee and Weinberg but they do not necessarily separate two different vacua: there are also unstable lumps that reach the same, symmetric vacuum, in the two asymptotic regions of the cylinder. On the physical side, these objects are ``anyon`` rings having both magnetic flux and electric charge. The difference with the topological vortices is that the rings are located along 1-cycles generating the homology of the cylinder, while the magnetic flux of the topological vortices is located along 1-cycles that are boundaries. This renders the reasons as to why analytic solutions exist on the cylinder intuitive. 

2. Computation of the dimension of the self-dual solution moduli space by means of an index theorem is pathological in the non-topological sector, because the magnetic flux is in general non-integer and the Witten and Atiyah-Singer indices do not coincide. We shall carefully analyze the index of the deformation operator of the self-dual solutions to show that the above mentioned pathology is due to the difference between the spectral densities of the continuous spectra of two supersymmetrically paired operators. In this way, we shall provide an explicit calculation of the correction to the heat kernel method that leads to the correct index in the non-topological sector of the CSH model, a matter previously justified only by indirect arguments \cite{bib: jlw}.

3. Finally, the attempt to interpret the number of degrees of freedom of the non-topological objects by comparing them with the exactly solvable solitons found in the gauged non-linear Schr\"{o}dinger equation has been not succesful, the reason being that in the fully relativistic theory the $|\phi|^6$ term, which we drop in the non-relativistic approximation, produces a non-ignorable force among solitons that changes the structure of the moduli space. We present an interpretation of the results of the index calculation, both on the plane and on the cylinder, by considering that only vortices can split apart from a general non-topological defect, so that the most general solution in the non-topological sectors cannot contain several non-topological solitons but rather only one of them surrounded by a number of vortices. Such an assembly can be obtained from a pure non-topological soliton with no cost in energy. 

The organization of the paper is as follows: in \S 2. we review the CSH model on the plane and try to clarify the division in topological sectors of the space of finite energy static configurations. Then, in \S 3., we deal with the same system on the cylinder, show that finding solitons is in this case equivalent to solving a conservative mechanical problem, and carry out the above index calculation. Section \S 4. is devoted to elaborating on the physical meaning of the non-topological defects; some brief final comments are included in \S 5.
\section{The CSH model on the plane. }
The Chern-Simons-Higgs model is defined by the action
\beq
S=\int d^3x\{\frac{\kappa}{4}\varepsilon^{\alpha\beta\gamma}A_\alpha F_{\beta\gamma}+\frac{1}{2}D_\mu\phi^* D^\mu\phi +U(\phi )\}\label{eq: 1}
\eeq
where $D_\mu\phi=\partial_\mu\phi +ieA_\mu\phi, \;  U(\phi )=\frac{\lambda}{8}|\phi |^2(|\phi |^2-v^2)^2$. We adopt natural units $\hbar =c=1$ and the space-time manifold is taken to be ${\bf M}^3({\bf R}^2)={\bf R}\times{\bf R}^2$ with metric $g_{\mu\nu}={\rm diag}(1, -1, -1)$. This action is unaffected by the gauge changes
\beq
A_\mu\rightarrow A_\mu -\partial_\mu\Lambda\; \; \; \; \; \; \; \phi\rightarrow e^{ie\Lambda}\phi
\eeq
The potential $U(\phi )$ reaches its minimun value for the constant configurations $\phi=0$ and $|\phi |=v$,
which are degenerate in energy. Quantization around the symmetric vacuum $\phi=0$ preserves the $U(1)$ symmetry. Expanding the action around $\phi =0$, one reads the particle spectrum from the quadratic terms: it is formed by two scalar bosons of mass  $m_\phi =\frac{\sqrt{\lambda}v^2}{2}$. Mass degeneration is due to $U(1)$ invariance but there are no photonic degrees of freedom because the gauge field propagation is governed by the pure Chern-Simons term. On quantizing around any of the asymmetric vacua $|\phi|=v$, however, the $U(1)$ symmetry is spontaneously broken. In the asymmetric phase, the Higgs mechanism combines with the Chern-Simons term to provide a mass for $A_\mu$; the vector field becomes massive by eating
one of the two scalar particles and the particle spectrum includes a polarized vector boson of mass $m_A=\frac{e^2v^2}{\kappa}$ and one scalar mode of mass $m_\phi =\sqrt{\lambda}v^2$. 

The Euler-Lagrange equations of the system that render the action functional (\ref{eq: 1}) extremal are
\begin{eqnarray}
\frac{\kappa}{2}\varepsilon^{\mu\alpha\beta}F_{\alpha\beta}&=&j^\mu\label{eq: 2}\\
D_\mu D^\mu\phi&=&-\frac{\lambda}{4}\phi (3|\phi |^2-v^2)(|\phi |^2-v^2)
\end{eqnarray}
where the conserved current is  $j^\mu =(\rho , \vec{j})={\displaystyle \frac{ie}{2}[\phi^*D^\mu\phi -\phi D^\mu\phi^*]}$. The time component of (\ref{eq: 2}), the so-called Chern-Simons-Gauss law, 
\beq
\kappa F_{12}=j^0\;\Rightarrow \; B=-\frac{\rho}{\kappa}\label{eq: 3}
\eeq
relates the electric charge and magnetic flux of the solutions of the theory and implies that a localized excitation of charge $Q$ must be considered as an anyon of fractional statistics ${\displaystyle \nu =\frac{Q^2}{2\pi\kappa}}$ . Equation (\ref{eq: 3}) can be stated in the form
\beq
A_0=\frac{\kappa B}{e^2|\phi |^2}-\frac{1}{e}\partial_0\arg\phi,\label{eq: 4}
\eeq
a constraint that eliminates $A_0$ as an independent degree of freedom. The energy-momentum tensor is
\beq
T^{\mu\nu}=\frac{1}{2}D^\mu\phi^*D^\nu\phi +\frac{1}{2}D^\nu\phi^*D^\mu\phi -g^{\mu\nu}[\frac{1}{2}D^\alpha\phi^*D_\alpha\phi -U(\phi )] 
\eeq
an expression in which there is not contribution from the Chern-Simons term due to its topological nature. In particular, for static configurations, the energy can be written as
\beq
E=\int d^2x\{  \frac{1}{4}\frac{\kappa^2}{e^2|\phi |^2}F_{ij}F^{ij}+\frac{1}{2}D_k\phi^*D_k\phi +U(\phi )\}\label{eq: 5}
\eeq
where the time-independent version of (\ref{eq: 4}) has been used. 

The configuration space ${\cal C}=\{\phi (x_1,x_2), A_i (x_1,x_2) / E < +\infty\}$ of the CSH-system is not topologically trivial. In the Higgs model the homotopical structure of ${\cal C}$ is determined from the homotopy of the set of zeroes ${\cal V}_\phi$ of $U$. In (2+1)-dimensions, the $i^{th}$ homotopy groups of ${\cal C}$
and ${\cal V}_\phi$ are related through the identity $\Pi_i ({\cal C})=\Pi_{i+1} ({\cal V}_\phi)$. Specially relevant is $\Pi_1 ({\cal V}_\phi)$ because it gives $\Pi_0 ({\cal C})$
and topological solitons arise when ${\cal C}$ is non-connected. Finite energy requires that
\beq
{\rm i)}\phi |_{\partial {\bf R}^2}\in{\cal V}_\phi,\; \; \; \; \; \; \; \; \; {\rm ii)} A|_{\partial {\bf R}^2}\in{\cal V}_A,\;\;\;\;\;\;\;\;\; {\rm iii)}D_k\phi |_{\partial {\bf R}^2}=0\label{eq: 6}
\eeq
at infinity. Here ${\cal V}_A$ is the set of pure gauges on the circle, ${\cal V}_A={\rm Maps} (S^1,{\bf R})$ if we take the radial gauge at long distances, and it is understood that the required values must be attained at the rate demanded for the integrability of (\ref{eq: 5}). When ${\cal V}_\phi=S^1$, the map meant by the Higgs field at infinity from $ \partial {\bf R}^2$ to
 ${\cal V}_\phi$ is the essential ingredient for stating that $\Pi_0 ({\cal C})=\Pi_1 ({\cal V}_\phi )={\bf Z}$.
Condition (\ref{eq: 6})-(iii) relates $A_i|_{\partial{\bf R}^2}$ to $\phi|_{\partial{\bf R}^2}$ and interpolation from $\partial {\bf R}^2$
towards the interior of ${\bf R}^2$  cannot change the boundary conditions (\ref{eq: 6}) without the cost of
infinite energy.

In the Chern-Simons Higgs Model things are more subtle because in this case the vacuum orbit is the disconnected union of a ``point", the symmetric vacuum, and a ``circle", the manifold of asymmetric vacua: ${\cal V}_\phi=\{\cdot\}\cup S^1$. Nevertheless, the boundary conditions (\ref{eq: 6}) also hold and from them we must derive the topological structure of ${\cal C}$. The main difference is that not only the first homotopy group of ${\cal V}_\phi$ is not trivial but also $\Pi_0 ({\cal V}_\phi)={\bf Z}_2$: ${\cal V}_\phi$ is disconnected itself. In three steps:

I. First, ${\cal C}$ splits into two disconnected sectors according to the asymptotic value of the Higgs field: ${\cal C}={\cal C}_0 \cup {\cal C}_v$. Continuous interpolation between Higgs configurations
such that $\phi|_{\partial{\bf R}^2}=0 \in {\cal C}_0$ or $|\phi||_{\partial{\bf R}^2}=v \in {\cal C}_v$ is not possible because they are associated to different elements in  $\Pi_0 ({\cal V}_\phi)$. Time-evolution from one sector to the 
other is thus forbidden and we use the topological charge $Q_1^T=|\phi ||_{\partial {\bf R}^2}$ to
label the two sub-spaces.

II. In both the CSH and the H models, condition (\ref{eq: 6}).(ii) allows a ``cohomological" interpretation of the topological classification of ${\cal C}$: The first Chern number ${\displaystyle Q_2^T=\frac{e\Phi_M}{2\pi}}$ associated with each configuration is well defined and finite in ${\cal C}$.
Through Stokes theorem, $Q_2^T$ is determined from (\ref{eq: 6}).(ii) and it is not necessarily an integer on a non-compact manifold as ${\bf R}^2$. Therefore, we define $Q_3^T=<Q_2^T>$, the fractionary part of $Q_2^T$, as a third topological label; $Q_3^T$ informs us about the holonomy of $A$ around the circle $\partial{\bf R}^2$.

III. In ${\cal C}_v$ the annihilation of the covariant derivative at infinity, (\ref{eq: 6}).(iii), implies trivial holonomy:
$Q_3^T=0$. Then, $Q_2^T \in {\bf Z}$ and it is a true topological charge which is equal to minus the winding number of the map $\phi : \partial{\bf R}^2 \longrightarrow S^1$. The ${\cal C}_v$ sector is then split into ${\bf Z}$ disconnected pieces. In the ${\cal C}_0$ sector, however, there are no requirements as regards the holonomy and the magnetic flux is not quantized: ${\cal C}_0$ is a connected (Hilbert) manifold.

 To summarize, the topological partition of ${\cal C}$ as
\begin{displaymath}
{\cal C}={\cal C}_0\cup_{n\in{\bf Z}}{\cal C}_v(n)
\end{displaymath}
with
\beq
\begin{array}{lll}Q_1^T[{\cal C}_0]=0,&Q_2^T[{\cal C}_0]\in{\bf R},&Q_3^T[{\cal C}_0]\in [0,1)\\Q_1^T[{\cal C}_v(n)]=v,&Q_2^T[{\cal C}_v(n)]=n,&Q_3^T[{\cal C}_v(n)]=0\end{array}
\eeq
differs from the analogous partition of the Higgs model configuration space by the existence of the
${\cal C}_0$ sector.

In order to search for soliton solutions in each sector  of ${\cal C}$ we write the energy functional (\ref{eq: 5}) in the form
\begin{displaymath}
E=E_{SD}\mp T
\end{displaymath}
where
\begin{eqnarray}
E_{SD}&=&\int d^2x\{\frac{1}{2}[\sqrt{G}F_{12}\pm\sqrt{2U}]^2+\frac{1}{2}|D_1\phi\pm iD_2\phi |^2\}\nonumber\\
T&=&\int d^2x\{\sqrt{2GU}F_{12}+\frac{i}{2}\varepsilon_{ij}D_i\phi^*D_j\phi\}
\end{eqnarray}
and ${\displaystyle G(|\phi |)=\frac{\kappa^2}{e^2|\phi |^2}}$. Choosing ${\displaystyle\lambda =\frac{e^4}{\kappa^2}}$, $T$ is proportional to the magnetic flux and
\beq
E=E_{SD}\mp\frac{1}{2}ev^2\int d^2xF_{12}
\eeq
hence we obtain the Bogomolnyi bound $E\geq\frac{1}{2}v^2|e\int d^2xF_{12}|$. Equality is attained if and only if $E_{SD}=0$, i.e.  if the self-duality equations
\begin{eqnarray}
F_{12}&\pm&\frac{e^3}{2\kappa^2}|\phi |^2(v^2-|\phi |^2)=0\nonumber\\
D_1\phi&\pm&iD_2\phi =0\label{eq: 9}
\end{eqnarray}
are satisfied. Note that, at the self-dual point, the scalar and vector bosons of the asymmetric phase have the same mass. Observe also that in the topological sectors with quantized magnetic flux, equations (\ref{eq: 9}) define absolute minima of the energy, which are stable solutions of the theory; in these sectors, the first-order equations imply the Euler-Lagrange ones. This implication is also true in the sector ${\cal C}_0$. 

Exact solutions of (\ref{eq: 9}) are not available. Radially symmetric self-dual solitons have been studied numerically and their main physical features described in \cite{bib: jlw} for any sector of ${\cal C}$. The existence of non-topological solitons also of radial type has been proved analytically in \cite{bib: sy} for the ${\cal C}_0$ sector. We review the main points of the analysis of Reference \cite{bib: sy} to compare with the results of \cite{bib: jlw} 

Using the ansatz
\beq
\phi (r, \theta )=vg(r)e^{in\theta }, \; \; \; \;  A_r=0, \; \; \; \;  A_\theta =\frac{a(r)-n}{e}\label{eq: 9x}
\eeq
and taking the upper sign in (\ref{eq: 9}), we obtain the system of first order equations
\begin{eqnarray}
\frac{1}{r}\frac{da}{dr}&=&\frac{m^2}{2}g^2(g^2-1)\nonumber\\
\frac{dg}{dr}&=&\frac{ag}{r}\label{eq: 10}
\end{eqnarray}
where ${\displaystyle m=m_A=m_\phi =\frac{e^2v^2}{\kappa}}$. The boundary conditions in ${\cal C}_s, s=0$ or $v$ are given by
\begin{eqnarray}
ng(0)=0&\;\;\;\;\;\;\;&a(0)=n\nonumber\\
vg(\infty)=s&\;\;\;\;\;\;\;&a(\infty)=-\alpha\label{eq: 10x}
\end{eqnarray}
where $\alpha=0$ in ${\cal C}_v$ and $\alpha >0$ in ${\cal C}_0$. The magnetic flux of the self-dual radial solitons is ${\displaystyle \Phi_M =\frac{2\pi}{e}(n+\alpha)}$. The electric charge and the energy are readily obtained from the Gauss law and the Bogomolny bound respectively:
\beq
Q_e=-\frac{2\pi}{e}\kappa(n+\alpha),\;\;\; E=\pi v^2(n+\alpha)
\eeq
In order to show the existence of solitons with these physical properties we change the variables to $t=\ln r$, $u=2\ln g$ and recast (\ref{eq: 10}) as
\beq
\frac{d^2u}{dt^2}=-e^{2t}\frac{dV}{du},\hspace{2.truecm}\frac{du}{dt}=2a\label{eq: mec}
\eeq
where ${\displaystyle V(u)=m^2e^u(1-\frac{e^u}{2})}$, see Fig.1. This is a non-conservative mechanical problem: a particle of mass unity and coordinate $u$ moving during the time interval $-\infty <t< +\infty$ under the influence of a time- and position- dependent force. The speed of the motion gives the vector field in (\ref{eq: 9x}) and hence the magnetic field. We now analyse this mechanical problem sector by sector:
\subsubsection*{Sector ${\cal C}_0$:} 
There are two possibilities which are addressed separately:
\begin{enumerate}
\item If $n\neq 0$ (non-topological vortices), the asymptotic conditions of the movement are
\begin{eqnarray}
u(-\infty)=-\infty,&\;\;\;\;\;\;\;\;\;&\dot{u}(-\infty)=2n\nonumber\\u(+\infty)=-\infty,&\;\;\;\;\;\;\;\;\;& \dot{u}(+\infty)=-2\alpha\label{eq: n19}
\end{eqnarray}
The particle comes from $-\infty$, is slowed down by the force until it reaches a maximun at some $u_{max} < 0$ (otherwise the conditions at $t=+\infty$ are impossible to meet) and then it comes back
to $-\infty$, see Fig.2. A continuous family of initial conditions $u(t_0)=u_{max}$, $\dot{u}(t_0)=0$, is compatible with the asymptotic behaviour (\ref{eq: n19}); thus  there is a family of non-topological vortices of vorticity $n$ parametrized by $u_{max}$. Because  the dynamics is not conservative, it is impossible to determine the function $\alpha =\alpha (u_{max})$ exactly. It is known that $\alpha$ is bounded from below by $n+2$ \cite{bib: jlw}, \cite{bib: sy}: multiplying the first equation in (\ref{eq: mec}) by $\dot{u}$ and integrating in $t$, we obtain
\begin{eqnarray}
\frac{1}{2}\dot{u}^2(+\infty )-\frac{1}{2}\dot{u}^2(-\infty )=-\int_{-\infty}^{+\infty}dte^{2t}\frac{dV}{dt}\Rightarrow\nonumber\\\Rightarrow 2\alpha^2-2n^2=-e^{2t}V|^{+\infty}_{-\infty}+2\int_{-\infty}^{+\infty}dte^{2t}V[u(t)] .
\end{eqnarray}
The boundary term vanishes: $0<\alpha<+\infty$, and hence, 
\beq
\lim_{t\rightarrow +\infty} e^{2t}V=\lim_{t\rightarrow +\infty}m^2e^{2t}e^{u(t)}=-\lim_{t\rightarrow +\infty}\frac{d^2u}{dt^2}=0.
\eeq

Since ${\displaystyle V(u)=\frac{m^2}{2}e^{2u}-e^{-2t}\frac{d^2u}{dt^2}}$, we deduce
\begin{eqnarray}
2\alpha^2-2n^2=-2\int_{-\infty}^{+\infty} dt\frac{d\dot{u}}{dt}+m^2\int_{-\infty}^{+\infty} dt e^{2t}e^{2u}\Rightarrow\nonumber\\\Rightarrow 2\alpha^2-2n^2-4\alpha-4n=m^2\int_{-\infty}^{+\infty} dt e^{2t}e^{2u}>0
\end{eqnarray}
and it follows that $\alpha >n+2$, as announced.
\item For $n=0$, the boundary conditions are
\begin{eqnarray}
u(-\infty)=u_0 <0,&\;\;\;\;\;\;\;\;\;&\dot{u}(-\infty)=0\nonumber\\ u(+\infty)=-\infty,&\;\;\;\;\;\;\;\;\;&\dot{u}(+\infty)=-2\alpha
\end{eqnarray}
The particle is pushed towards $-\infty$ from the very beginning. There is a family of non-topological solitons parametrized by $u_0\in (-\infty,0)$ and the lower bound for $\alpha$ is now simply 2.
\end{enumerate}
\subsubsection*{Sector ${\cal C}_v(n)$:} If $n \neq 0$, the requirements
\begin{eqnarray}
u(-\infty)=-\infty,&\;\;\;\;\;\;\;\;\;&\dot{u}(-\infty)=2n\nonumber\\ u(+\infty)=0,&\;\;\;\;\;\;\;\;\;&\dot{u}(+\infty)=0
\end{eqnarray}
show that the ``topological vortices'' appear as separatrices between the non-topological ones and the infinite energy solutions, see Fig.2. If $n=0$, we conclude that the only solution is the asymmetric vacuum, given by the equilibrium constant configuration $u(t)=0$. The topological vortices are the Chern-Simons peers of the Nielsen-Olesen vortices in the AHM. However, due to the $|\phi|^2$ factor in front of the standard Higgs potential, the magnetic flux is distributed over an annulus centred at the zero of order $n$ 
of the Higgs field, instead of being concentrated on a disk.

Apart from the above configurations, corresponding to defects centered on the origin of the plane, system (\ref{eq: 9}) should have other solutions in ${\cal C}$ that describe assemblies of such vortices or solitons. In particular, the dimension of the moduli space ${\cal M}$, the quotient by the gauge group of the set of solutions of the self-duality equations, has been calculated in \cite{bib: jlw} by index theorem techniques finding that: 
\begin{itemize}
\item Sector ${\cal C}_v(n)$:  ${\displaystyle \, \dim{\cal M}=2[\frac{e\Phi_M}{2\pi}]=2n}$
\item Sector ${\cal C}_0$:  ${\displaystyle \, \dim{\cal M}=2[\frac{e\Phi_M}{2\pi}]-2=2[\alpha ]+2n-2}$
\end{itemize}
Proof of the existence of topological vortex solutions of the CSH model with arbitrary location of the vortex centres has been provided in Reference \cite{bib: wg}. We finish this section by noting that the authors of \cite{bib: jlw} have also elaborated on domain wall solutions. They are symmetric along one direction of the plane and do not belong to ${\cal C}$.
\section{The CSH model on the cylinder. }
In this section we consider the CSH theory defined on a cylinder, i.e. the spatial manifold is $C={\bf R}\times S^1$ with coordinates $(x^1,x^2)$ such that $x^1\in (-\infty , +\infty )$, $x^2\in (0, 2\pi )$. The space-time manifold is ${\bf M}_3(C)={\bf R}\times C$ equipped with the Minkowskian metric of Section \S 2. Although one of the coordinates becomes periodic, in the energy functional  (\ref{eq: 5}) for static configurations only the integration domain changes from ${\bf R}^2$ to
$C$. There is no difference in the Bogomolny splitting and the solutions of the first order equations (\ref{eq: 9}) solve the Euler-Lagrange equations.

On the cylinder, the topological structure of the configuration space ${\cal C}$ is richer than on the plane.  Finite energy now requires
\beq
{\rm i)}\phi |_{\partial C}\in {\cal V}_\phi,\; \; \; \; \; \; \; \; \; \; \; {\rm ii)}A|_{\partial C}\in{\cal V}_A,\;\;\;\;\;\;\;\;\;\;\; {\rm iii)}D_k\phi |_{\partial C}=0\label{eq: 11}
\eeq
where $\partial C$ is the boundary of the cylinder; it is the disconnected union of two circles $\partial C=\partial C_+\cup\partial C_-$, $\partial C_\pm=\{(\pm\infty, x_2)/0\leq x_2\leq 2\pi\}=S_{\pm}^1$ and hence ${\cal V}_A={\rm Maps} (S^1\cup S^1,{\bf R})$. Both $\partial C$ and ${\cal V}_\phi$ are disconnected: $\Pi_0({\cal V}_\phi)={\bf Z}_2$, $\Pi_0(\partial C)={\bf Z}_2$. We describe the topological structure of ${\cal C}$ in comparison to the topological partition of Section \S 2.

I. First, ${\cal C}$ is split into four separate sectors according to the values reached by the Higgs field at $\partial C_-(\partial C_+)$: ${\cal C}={\cal C}_0^0\cup{\cal C}_0^v\cup{\cal C}_v^0\cup{\cal C}_v^v$. The subindices (superindices) in each Sector refer to the asymptotics of $|\phi|$ at $-\infty$ ($+\infty$). There is no continuous evolution between the four types of behaviour allowed by condition (\ref{eq: 11}) (i), and the conservation of the topological charge $Q^T_1 =|\phi ||_{\partial C_+}-|\phi ||_{\partial C_+}$ is ensured.

II.The first Chern number,
\bdm
Q_2^T=\frac{e\Phi_M}{2\pi}=\frac{1}{2\pi}\int_{\partial C_-}eA_2dx_2-\frac{1}{2\pi}\int_{\partial C_+}eA_2dx_2
\edm
which provides a finer classification of ${\cal C}$ related to the condition (\ref{eq: 11}) (ii), differs slightly from 
its analogue on the plane.
$Q_3^T=<Q_2^T>$ measures the difference between the holonomies of the $U(1)$-connection over the two asymptotic circles (1-cycles which are not boundaries in the homological sense).

III. Nullity of the covariant derivative at infinity, condition (\ref{eq: 11}).(iii), forces trivial holonomy only in ${\cal C}_v^v$. In this sector, $Q_2^T$ is an integer-conserved charge that is equal to the difference between the winding numbers of $\phi$ on $\partial C_\pm$ and induces the splitting of the ${\cal C}_v^v$ sector
into ${\bf Z}$ disconnected pieces: ${\cal C}_v^v=\cup_{n\in{\bf Z}}{\cal C}_v^v(n)$.

The topological structure of the configuration space is then:
\begin{displaymath}
{\cal C}={\cal C}_0^0\cup{\cal C}_0^v\cup{\cal C}_v^0\cup_{n\in{\bf Z}}{\cal C}_v^v(n)
\end{displaymath}
such that,
\beq
\begin{array}{llll}
Q_1^T[{\cal C}_0^0]=0,&Q_1^T[{\cal C}_0^v]=v,&Q_1^T[{\cal C}_v^0]=-v,&Q_1^T[{\cal C}_v^v(n)]=0\nonumber\\Q_2^T[{\cal C}_0^0]\in{\bf R},&Q_2^T[{\cal C}_0^v]\in{\bf R},&Q_2^T[{\cal C}_v^0]\in{\bf R},&Q_2^T[{\cal C}_v^v(n)]=n\\Q_3^T[{\cal C}_0^0]\in [0,1),&Q_3^T[{\cal C}_0^v]\in [0,1),&Q_3^T[{\cal C}_v^0]\in [0,1),&Q_3^T[{\cal C}_v^v(n)]=0 \end{array}
\eeq

The self-duality equations of the abelian Higgs model can be defined on any (compact or not) Riemann surface, see \cite {bib:br}. It has been shown by Bradlow that solutions exist in the compact case provided that a certain inequality between the surface area and the soliton magnetic flux is satisfied. On a non-compact surface, such as a infinite cylinder this is automatic; there are vortex solutions that can be obtained by patching together the topological vortices arising in each of the two charts isomorphic to ${\bf R}^2$. It seems plausible that topological vortices of Chern-Simons type living on a cylinder in ${\cal C}_v^v (n)$
can be constructed in a similar manner from their cousins on the plane that belong to ${\cal C}_v (n)$.
The conjecture is supported by the fact that CSH topological vortices in ${\bf R}^2$ reach the vacuum value at a similar rate to that of the self-dual vortices in the Higgs model. It is less clear, however, whether non-topological vortices can be extended to a cylinder; we shall comment on this possibility in Section \S 4 after analyzing the local structure of the moduli space.
\subsection{The axial solitons. }
The main interest in studying the CSH system on a cylinder is that in the Sectors ${\cal C}_0^v$, ${\cal C}_v^0$, not arising on the plane, it is posssible to find exact axially symmetric self-dual solutions analytically; these solutions correspond to the domain walls of Jackiw et al., although on a cylinder their
energy is finite. Moreover, similar solutions in the ${\cal C}_0^0$ Sector also exist, and are the counterpart of the non-topological solitons on the plane. Exact solutions of the vortex equations are very rare; the other cases where they are known are the Higgs model on the half-plane with the Poincar\'e metric
and the non-relativistic vortices of Jackiw-Pi, \cite{bib: jp}. In those cases, the self-duality equations become equivalent to the (solvable)  Liouville equation.

Searching for axially symmetric solutions of (\ref{eq: 9}), we consider the ansatz
\beq
\phi (x_1, x_2)=vg(x_1)e^{ikx_2 }, \; \;  eA_1=0, \; \;  eA_2=a(x_1)-k , \; \;  k \in {\bf Z},
\eeq
the $k^{th}$ term of the Fourier expansion in $S^1$, suggested by the symmetry. In a $U(1)$ gauge theory on a cylinder there are bona fide gauge transformations given by $\Lambda (x_1, x_2)=-\omega x_2$ if and only if $\omega \in {\bf Z}$. The ansatz can be simplified to the gauge equivalent form:
\beq
\phi (x_1, x_2)=vg(x_1), \; \; \; \;  eA_1=0, \; \; \; \;  eA_2 =a(x_1)\label{eq: 12}
\eeq
Choosing the upper sign in (\ref{eq: 9}) and plugging ansatz (\ref{eq: 12}) in the self-duality equations, we find
\begin{eqnarray}
\frac{da}{dx_1}&=&\frac{m^2}{2}g^2(g^2-1)\nonumber\\
\frac{dg}{dx_1}&=&ag\label{eq: 13}
\end{eqnarray}
In the Sector ${\cal C}_r^s$  the axial Bogomolnyi equations must be solved together with the boundary conditions
\begin{eqnarray}
vg(-\infty)=r&\;\;\;\;\;\;&a(-\infty)=\alpha\nonumber\\
vg(+\infty)=s&\;\;\;\;\;\;&a(+\infty)=-\beta   .\label{eq: cond}
\end{eqnarray}
Here, $\alpha=0$ in ${\cal C}_v^v(0)$ and ${\cal C}_v^0$, $\beta=0$ in ${\cal C}_v^v(0)$ and ${\cal C}_0^v$, and $\alpha,\beta >0$ otherwise. The magnetic flux, electric charge and energy of an axial self-dual solution are
\beq
\Phi_M=\frac{2\pi}{e}(\alpha +\beta),\;\;\;\; Q=-\frac{2\pi}{e}\kappa(\alpha +\beta),\;\;\;\; E=\pi v^2(\alpha +\beta)\label{eq: cond2}
\eeq
Changing variables to  $t=x_1$ and $u=2\ln g$, (\ref{eq: 13}) becomes:
\beq
\frac{d^2u}{dt^2}=-\frac{dV}{du}, \hspace{2.truecm}  \frac{du}{dt}=2a\label{eq: n31}
\eeq
Again we encounter the same $V(u)$ as in the mechanical problem of Section \S 2. The big difference is that now the system (\ref{eq: n31}) is conservative, and hence, solvable. The mechanical energy is
\beq
{\cal E}=\frac{1}{2}(\frac{du}{dt})^2+V(u)
\eeq
The geometric reason for having a time-independent force in this case can be traced back to the fact that the cylinder and the plane are not isometric but only conformally equivalent; the conformal factor must be compensated on the plane making the vortex equations much harder to solve. Thus, axial self-dual solutions are in one-to-one correspondence with the solutions of the differential equation
\beq
\frac{1}{2}(\frac{du}{dt})^2={\cal E}-m^2e^u(1-\frac{e^u}{2})\label{eq: 14}
\eeq
which must be integrated using the boundary conditions appropriate to  each sector: 
\subsubsection*{Sector ${\cal C}_v^v(0)$: } 
In this Sector the asymptotic conditions are
\beq
\begin{array}{ccc}u(-\infty )=0&\; \; \; \; \; \; &\dot{u}(-\infty )=0\\u(+\infty )=0&\; \; \; \; \; \; &\dot{u}(+\infty )=0\end{array}
\eeq
and by inspection of Fig.1 we conclude that the only possible solution is the equilibrium $u(t)=0$. The corresponding field configuration is
\beq
\phi(x_1, x_2)=v\; \; \; \; \; \;  eA_2(x_1, x_2)=0
\eeq
i.e. the only solution is the asymmetric vacuum. In the other ${\cal C}_v^v(n)$ sectors, there are not regular radially symmetric solutions because they would imply that $\phi(x_1,x_2)=0$ for some finite $x_1$ and all $x_2$, but the self-duality equations require discreteness of the zeroes of $\phi$ \cite{bib: wg}.
\subsubsection*{Sector ${\cal C}_0^v$: }
Here, the asymptotic conditions are
\beq
\begin{array}{ccc}u(-\infty )=-\infty&\; \; \; \; \; \; &\dot{u}(-\infty )=2\alpha\\u(+\infty )=0&\; \; \; \; \; \; &\dot{u}(+\infty )=0\end{array}
\eeq
and the movement is represented by the upper line of Fig.3. Conservation of energy implies
\beq
{\cal E}=\frac{1}{2}(2\alpha )^2=\frac{m^2}{2}\ \Rightarrow \ \alpha =\frac{m}{2}
\eeq
Equation (\ref{eq: 14}) is easily integrated to give $u(t)=-\ln (1+Ke^{-mt})$, where $K$ is a positive integration constant. The particular value $K=1$ gives the soliton centered on $x_1=0$. The associated field configuration and magnetic field of the axial solutions are
\begin{eqnarray}
\phi(x_1,x_2)&=&\frac{v}{\sqrt{2}}e^{\frac{mx_1}{4}}{\rm sech}^{\frac{1}{2}}(\frac{m}{2}x_1)\nonumber\\
eA_2(x_1,x_2)&=&\frac{m}{4}e^{-\frac{mx_1}{2}}{\rm sech}(\frac{m}{2}x_1)\nonumber\\
eB(x_1,x_2)&=&\frac{m^2}{8}{\rm sech}^2(\frac{m}{2}x_1)
\end{eqnarray}
The stability of this solution against decay to the vacuum is ensured by the conservation of the topological charge $Q_1^T$, even though the magnetic flux is not quantized
\subsubsection*{Sector ${\cal C}_v^0$: }
The boundary conditions are obtained from those in ${\cal C}_0^v$ simply by exchanging the behaviours at $\pm\infty$. A solution in this sector is thus the parity transform of any solution in ${\cal C}_0^v$:
\beq
\phi (x_1, x_2)\rightarrow \phi (-x_1, x_2), \; \; \; \;  A_2(x_1, x_2)\rightarrow  -A_2(-x_1, x_2)
\eeq
\subsubsection*{Sector ${\cal C}_0^0$. }
In this sector the movement is required to satisfy
\beq
\begin{array}{ccc}u(-\infty )=-\infty&\; \; \; \; \; \; &\dot{u}(-\infty )=2\alpha\\u(+\infty )=-\infty&\; \; \; \; \; \; &\dot{u}(+\infty )=-2\beta\end{array}
\eeq
and corresponds to the lower line in Fig.3. From energy conservation
\beq
{\cal E}=\frac{1}{2}(2\alpha )^2=\frac{1}{2}(2\beta )^2\ \Rightarrow\ \alpha =\beta ,
\eeq
and we expect a family of solutions parametrized by $\alpha\in (0, \frac{m}{2})$. The turning point occurs at $u_0=\ln [1-\sqrt{1-\frac{4\alpha^2}{m^2}}]$ and, assuming that it is reached when $t=0$, equation (\ref{eq: 14}) implies
\beq
|t|=I(u_0)-I(u)\label{eq: I},
\eeq
where
\beq
{\displaystyle I(w)=\int\frac{dw}{\sqrt{4\alpha^2-2m^2e^w(1-\frac{e^w}{2})}}}.
\eeq
Inverting (\ref{eq: I}) one has
\beq
u(t)=-\ln \{\frac{m^4}{\alpha^2}[1+\sqrt{1-\frac{4\alpha^2}{m^2}}\cosh (2\alpha |t|)]\};
\eeq
The field configurations and magnetic field of a self-dual non-topological soliton of the CSH system on a cylinder are:
\begin{eqnarray}
\phi (x_1, x_2)&=&\frac{\alpha v}{m^2}\{1+\sqrt{1-\frac{4\alpha^2}{m^2}}\cosh (2\alpha x_1)\}^{-\frac{1}{2}}\nonumber\\eA_2(x_1, x_2)&=&-\frac{\alpha\sqrt{1-\frac{4\alpha^2}{m^2}}{\rm sinh} (2\alpha x_1)}{1+\sqrt{1-\frac{4\alpha^2}{m^2}}\cosh(2\alpha x_1)}\\
eB(x_1, x_2)&=&2\alpha^2\sqrt{1-\frac{4\alpha^2}{m^2}}\frac{\cosh(2\alpha x_1)-\sqrt{1-\frac{4\alpha^2}{m^2}}}{[1+\sqrt{1-\frac{4\alpha^2}{m^2}}\cosh(2\alpha x_1)]^2}\nonumber
\end{eqnarray}
Note that because in this sector $Q_1^T=0$ and $Q_3^T$ is arbitrary, there is no topological reason to expect stability.
\subsection{The dimension of the moduli space. }
We now address the computation of the dimension of the moduli space of self-dual CSH solitons on a cylinder as the index of the deformation operator \cite{bib: jlw}, an elliptic differential operator defined as follows: given a solution $\Gamma =(\phi, eA )$ of the Bogomolnyi equation, we obtain another solution $\Gamma^\prime =\Gamma +\delta\Gamma$ if the $L^2$-normalizable small deformation $\delta\Gamma$ satisfies the linearization of (\ref{eq: 9}), i.e. 
\begin{eqnarray}
(\partial_1-eA_2)\delta\phi_1+(-\partial_2-eA_1)\delta\phi_1-e\phi_2\delta A_1-e\phi_1\delta A_2&=&0\nonumber\\(\partial_2+eA_1)\delta\phi_1+(\partial_1-eA_2)\delta\phi_2+e\phi_1\delta A_1-e\phi_2\delta A_2&=&0\nonumber\\\frac{m^2}{v^2}\phi_1[1-2\frac{|\phi |^2}{v^2}]\delta\phi_1+\frac{m^2}{v^2}\phi_2[1-2\frac{|\phi |^2}{v^2}]\delta\phi_2-e\partial_2\delta A_1+e\partial_1\delta A_2&=&0\label{eq: 15}
\end{eqnarray}
where $\phi =\phi_1+i\phi_2$, $\delta\phi =\delta\phi_1+\delta\phi_2$. To avoid deformations corresponding to gauge transformations, we choose to work in the Coulomb gauge
\beq
\partial_1\delta A_1+\partial_2\delta A_2=0\label{eq: 16}
\eeq
This condition eliminates all the gauge modes in the sectors ${\cal C}^v_v(n), {\cal C}_0^v$ and ${\cal C}_v^0$, although a mode in ${\cal C}_0^0$ corresponding to a global gauge transformation is not supressed; we shall subtract its contribution at the end of the computation.

From (\ref{eq: 15}), (\ref{eq: 16}) we see that the dimension of the moduli space is equal to the dimension of the kernel of the elliptic differential operator
\beq
{\cal D}=\left(\begin{array}{cccc}\partial_1-eA_2&-\partial_2-eA_1&-\phi_2&-\phi_1\\\partial_2-eA_1&\partial_1-eA_2&\phi_1&-\phi_2\\\frac{m^2}{v^2}\phi_1[1-2\frac{|\phi |^2}{v^2}]&\frac{m^2}{v^2}\phi_2[1-2\frac{|\phi |^2}{v^2}]&-\partial_2&\partial_1\\0&0&-\partial_1&-\partial_2\end{array}\right)
\eeq
The index of the deformation operator is defined by ${\rm ind}{\cal D}=\dim\ker{\cal D}-\dim\ker{\cal D}^\dagger$; in particular ${\rm ind}{\cal D}=\dim\ker{\cal D}$ if $\ker{\cal D}^\dagger =\{ 0\}$. It is not difficult to convince ourselves that this is indeed the case. The strategy is as follows: introduce ${\cal R}={\cal TD}^\dagger$, where
\beq
{\cal T} =\left(\begin{array}{cccc}\phi_2&-\phi_1&-\partial_1&-\partial_2\\\phi_1&\phi_2&-\partial_2&-\partial_1\\0&0&-\phi_2&-\phi_1\\0&0&-\phi_1&-\phi_2\end{array}\right)
\eeq
is an operator such that $\ker{\cal T}=\{ 0\}$. Then $\ker{\cal R}=\ker{\cal D}^\dagger$ and by explicit computation it is immediate to see that ${\cal R}$ is a second-order differential operator without $L^2$-normalizable zero eigenfunctions provided that a second-order ordinary differential equation arising in the process has, generically, no normalizable solution.

Therefore, $\dim\ker{\cal D}^\dagger =0$ and the dimension of the moduli space ${\cal M}$ is given by the index of ${\cal D}$, which can be computed by means of the asymptotic expansion of the kernel of the ``heat equation''. We write ${\cal D}$ and ${\cal D}^\dagger$ as a Dirac operator and its conjugate
\begin{eqnarray}
{\cal D}&=&i\alpha_1\partial_1+i\alpha_2\partial_2+Q(x)\nonumber\\{\cal D}^\dagger&=&i\alpha_1^\dagger\partial_1+i\alpha_2^\dagger\partial_2+Q(x)^\dagger   .\label{eq: 17}
\end{eqnarray}
The $\alpha_k$ matrices satisfy the Clifford algebra
\beq
\alpha_i^\dagger\alpha_j+\alpha_j^\dagger\alpha_i=\alpha_i\alpha_j^\dagger +\alpha_j\alpha_i^\dagger =2\delta_{ij}
\eeq
and the differential operators $W_+={\cal D}^\dagger{\cal D}$ and $W_-={\cal D}{\cal D}^\dagger$ are of the form $W_\pm =-\nabla^2+L_\pm$, where $L_\pm$ are defined by
\begin{eqnarray}
L_-&=&i[\alpha_iQ^\dagger +Q\alpha_i^\dagger ]\partial_i+i\alpha_i(\partial_iQ^\dagger )+QQ^\dagger\nonumber\\L_+&=&i[\alpha_i^\dagger Q +Q^\dagger\alpha_i ]\partial_i+i\alpha_i^\dagger (\partial_iQ)+Q^\dagger Q
\end{eqnarray}
The kernels of the heat equations associated with $W_\pm$ are $h_{W_\pm}(t)={\rm Tr}e^{-tW_\pm}$ and because $W_+$ and $W_-$ share the non-zero eigenvalues
\beq
{\rm ind}{\cal D}=h_{W_+}(t)-h_{W_-}(t)+\int d\lambda\int d^2x[(\psi_\lambda^-(x), \psi_\lambda^-(x))-(\psi_\lambda^+(x), \psi_\lambda^+(x))]e^{-t\lambda^2}\label{eq: apar}
\eeq
where $\psi_\lambda^\pm$ are the eigenfunctions of $W_\pm$, $W_\pm\psi_\lambda^\pm =\lambda^2\psi_\lambda^\pm$; they are related through the equation  $\psi_\lambda^-=\frac{1}{\lambda}{\cal D}\psi_\lambda^+$. The last term in (\ref{eq: apar}) represents the difference between the spectral densities of the continuous spectra of $W_\pm$. It must be added to the supertrace $h_{W_+}(t)-h_{W_-}(t)$ because the operator that implements the transformation between $\psi_\lambda^\pm$ includes derivatives and therefore changes the spectral density. This is the key point that will lead us to obtain the exact index instead of the non-integer results of reference \cite{bib: jlw}. By introducing the explicit form (\ref{eq: 17}) of ${\cal D}$ it is easy to see that \cite{bib: stn}
\beq
(\psi_\lambda^-(x), \psi_\lambda^-(x))-(\psi_\lambda^+(x), \psi_\lambda^+(x))=-\frac{i}{\lambda^2}\partial_k(\psi_\lambda^+(x), \alpha_k^\dagger{\cal D}\psi_\lambda^+(x))
\eeq
and we then have
\begin{eqnarray}
{\rm ind}{\cal D}&=&h_{W_+}(t)-h_{W_-}(t)+\Delta(t)\\\Delta(t)&=&\int d\lambda\int_0^{2\pi}dx_2(\psi_\lambda^+, -i\alpha_1^\dagger\psi_\lambda^+)\frac{e^{-t\lambda^2}}{\lambda^2}|^{x_1=+\infty}_{x_1=-\infty}  .\label{eq: ind}
\end{eqnarray}
Bearing in mind that ${\rm ind}{\cal D}$ is independent of $t$, we can choose the limit $t\rightarrow 0$ and use the corresponding asymptotic expressions of $h_{W_\pm}(t)$; namely:
\begin{eqnarray}
e^{-tW_+}&\simeq& e^{-t\nabla^2}[1-tL_+ +o(t^2)]\nonumber\\e^{-tW_-}&\simeq &e^{-t\nabla^2}[1-tL_-+o(t^2)]   .
\end{eqnarray}
Then,
\beq
h_{W_+}(t)-h_{W_-}(t)\simeq t\int_Cd^2x{\rm tr}\langle x|e^{-t\nabla^2}(L_--L_+)|x\rangle
\eeq
when $t\rightarrow 0$ up to the first order in the $t$ expansion, and 
\begin{eqnarray}
h_{W_+}(t)-h_{W_-}(t)&\simeq& t\int_Cd^2x\int_Cd^2y\langle x|e^{-t\nabla^2}|y\rangle{\rm tr}\langle y|(L_--L_+)|x\rangle =\nonumber\\&=&it\int_Cd^2x\langle x|e^{-t\nabla^2}|x\rangle{\rm tr}F(x)
\end{eqnarray}
where
\beq
F(x)=-i\langle x|L_--L_+|x\rangle=\alpha_i\partial_iQ(x)^\dagger-\alpha_i^\dagger\partial_iQ (x)
\eeq
and then, using the explicit form of that matrices
\beq
{\rm tr}F(x)=4ieF_{12}(x)
\eeq
In the cylinder, the spectrum of the Laplacian is
\begin{eqnarray}
\nabla^2|k, n\rangle&=&-(k^2+n^2)|k, n\rangle\nonumber\\\langle x_1, x_2|k, n\rangle&=& \frac{1}{2\pi}e^{i(kx_1+nx_2)}\; \; \; \; \; \; k\in{\bf R}, n\in{\bf Z}
\end{eqnarray}
and thus
\beq
h_{W_+}(0)-h_{W_-}(0)=\lim_{t\rightarrow 0}\frac{it}{(2\pi )^2}[\sum_{n\in{\bf Z}}e^{-tn^2}\int_{-\infty}^{+\infty}dke^{-tk^2}]\int_Cd^2x {\rm tr}F(x)
\eeq
For a small enough value of $t$, it is possible to substitute the summatory by an integral and therefore
\beq
{\rm ind}{\cal D}=h_{W_+}(0)-h_{W_-}(0)+\Delta=\frac{e\Phi_M}{\pi}+\Delta,
\eeq
$\Delta\equiv\lim_{t\rightarrow 0}\Delta(t)$. To evaluate this correction it is convenient to rewrite (\ref{eq: ind}) as
\beq
\Delta(t)=\int d\lambda\int_0^{2\pi}dx_2(\psi_\lambda^+, -i\alpha_1^\dagger{\cal D}\frac{e^{-W_+}}{W_+}\psi_\lambda^+)|_{x_1=-\infty}^{x_1=+\infty}
\eeq
but in the asymptotic regions of the cylinder the autofunctions have the form
\beq
\psi_\lambda^+=\frac{V_\lambda}{2\pi}e^{ i(kx_1+nx_2)},\ \ \ \ V_\lambda^\dagger V_\lambda=1
\eeq
where $V_\lambda$ is a constant column vector; we can then replace the differential operators by matrices
\beq
W_+\psi_\lambda^+=M(k,n)\psi_\lambda^+,\ \ \ \ -i\alpha_1^\dagger{\cal D}\psi_\lambda^+=S(k,n)\psi_\lambda^+
\eeq
where
\begin{eqnarray}
S(k,n)&=&ik+i\alpha_1^\dagger\alpha_2n-i\alpha_1^\dagger Q\nonumber\\
M(k,n)&=&k^2+n^2-(\alpha_1^\dagger Q+Q^\dagger\alpha_1)k-(\alpha_2^\dagger Q+Q^\dagger\alpha_2)n+Q^\dagger Q
\end{eqnarray}
such that
\beq
\Delta=\sum_{n\in{\bf Z}}\int_{-\infty}^{+\infty}\frac{dk}{2\pi}{\rm tr}\{S(k,n)M^{-1}(k,n)\}|_{x_1=-\infty}^{x_1=+\infty}
\eeq
After some straightforward but tedious work, the trace can be computed and the results are as follows:
\begin{enumerate}
\item Topological boundary conditions: If the fields on one border of the cylinder are
\beq
\phi=v,\ \ \ \ \ \ eA_2=0
\eeq
one obtains
\beq
{\rm tr}\{S(k,n)M^{-1}(k,n)\}=kH(k^2,n^2)
\eeq
and the integration in $k$ renders the contribution of this border to $\Delta$ null.
\item Non-topological boundary conditions: If, by contrast, the asymptotic fields are
\beq
\phi=0,\ \ \ \ \ \ eA_2\neq 0
\eeq
the trace gives
\beq
{\rm tr}\{S(k,n)M^{-1}(k,n)\}=kG(k^2,n^2)-\frac{2eA_2(k^2-n^2+e^2A_2^2)-4n^2eA_2}{(k^2+n^2+e^2A_2^2)^2-4n^2e^2A_2^2}
\eeq
and $\Delta$ picks a contribution
\beq
\pm Z=\pm\sum_{n\in{\bf Z}}\int_{-\infty}^{+\infty}\frac{dk}{2\pi}\frac{2eA_2(k^2-n^2+e^2A_2^2)-4n^2eA_2}{(k^2+n^2+e^2A_2^2)^2-4n^2e^2A_2^2}
\eeq
where the plus (minus) sign must be used if we are at the $x_1=-\infty$ ($x_1=+\infty$) border of the cylinder. The integral and the infinite summatory are not difficult to calculate (for the latter we use the standard regularization, see \cite{bib: ang})
\begin{eqnarray}
Z&=&\frac{1}{2\pi}\sum_{n\in{\bf Z}}\{{\rm arctg}(\frac{k}{eA_2+n})+{\rm arctg}(\frac{k}{eA_2-n})\}|_{k=-\infty}^{k=+\infty}=\nonumber\\ &=&\frac{1}{2}\sum_{n\in{\bf Z}}\{{\rm sgn}(eA_2+n)+{\rm sgn}(eA_2-n)\}=1-2<eA_2>
\end{eqnarray}
\end{enumerate}
The outcome of the above treatment is therefore the formula
\begin{eqnarray}
{\rm ind}{\cal D}&=&\frac{e\Phi_M}{\pi}+\delta_{0,\phi(\partial C_-)}(1-2<eA_2(\partial C_-)>)-\nonumber\\ &-&\delta_{0,\phi(\partial C_+)}(1-2<eA_2(\partial C_+)>)
\end{eqnarray}
whose application to each sector is as follows:
\subsubsection*{Sector ${\cal C}_v^v(n)$:}
Here, the correction $\Delta$ vanishes: the boundary conditions allow one to compactify the cylinder and therefore the index is the same as over a compact manifold
\beq
{\rm ind}{\cal D}=2\frac{e\Phi_M}{2\pi}=2n
\eeq
\subsubsection*{Sector ${\cal C}_0^v$:}
In this sector, the boundary conditions (\ref{eq: cond}) and the flux (\ref{eq: cond2}) give
\beq
{\rm ind}{\cal D}=2\alpha+(1-2<\alpha>)=2[\alpha]+1=2[\frac{e\Phi_M}{2\pi}]+1
\eeq
because $\beta=0$.
The same result also applies to the sector ${\cal C}_v^0$.
\subsubsection*{Sector ${\cal C}_0^0$:}
Here, $\alpha=\beta$ in (\ref{eq: cond}). Then, because $<-\alpha>=1-<\alpha>$
\beq
{\rm ind}{\cal D}=4\alpha+(1-2<\alpha>)-(1-2<-\alpha>)=4[\alpha]+2
\eeq\\
Having finished the index computation, and recalling the spurious global gauge mode from the ${\cal C}_0^0$ sector, we summarize the results obtained for the dimensions of the moduli spaces of the self-dual solutions in the distinct sectors:
\begin{itemize}
\item Sector ${\cal C}_v^v(n)$: \ \ \ ${\displaystyle \dim{\cal M}=2\frac{e\Phi_M}{2\pi}=2n}$
\item Sectors ${\cal C}_0^v$ and ${\cal C}_v^0$: \ \ \ ${\displaystyle \dim{\cal M}=2[\frac{e\Phi_M}{2\pi}]+1=2[\frac{m}{2}]+1}$
\item Sector ${\cal C}_0^0$: $\ \ \ \dim{\cal M}=2[eA_2(-\infty, x_2)]-2[eA_2(+\infty, x_2)]-1=4[{\displaystyle \frac{e\Phi_M}{2\pi}}]+1$
\end{itemize}
To provide a physical interpretation for these results as well as for those previously seen on the plane is the subject of the next section.
\section{On non-topological solitons. }
The structure of the moduli space of self-dual solutions in the ${\cal C}_v (n)$ sector of the CSH model on ${\bf R}^2$  has been thoroughly analysed by Wang \cite{bib: wg}. Following the fundamental work of Jaffe and Taubes \cite{bib: jt} on the AHM, he found that the dimensionality of the moduli space corresponds to the $2n$ translational degrees of freedom of $n$ non-interacting vortices whose centers coincide with the zeroes of the Higgs field on the plane. To our knowledge there is no analogous analysis of the physical meaning of the moduli space of the non-topological vortices in the literature. In this section we tackle the subject bearing in mind that comparison of the results for the non-topological solitons and vortices on the cylinder reveals that the latter are mixed objects formed by a basic non-topological soliton plus several vortices. Therefore, a physical interpretation of the non-topological soliton moduli space suffices to understand all the non-topological defects. 

We write ${\displaystyle \phi =e^{\frac{u+i\Theta}{2}}}$ and set
\begin{eqnarray*}
Z(\phi )&=&\{{\rm zeroes\;  of\;  }\phi{\rm\;  en\;  }{\bf R}^2\}\nonumber\\S(\Theta )&=&\{{\rm singular\;  points\;  of\; } \Theta\;  {\rm en\; }{\bf R}^2\} . \nonumber
\end{eqnarray*}
The analysis carried out in \cite{bib: jt} shows that for self-dual configurations $Z(\phi )$ is discrete, $S(\Theta )=Z(\phi )$, and near each $\vec{x}_k\in Z(\phi )$
\beq
\phi(\vec{x}) \simeq (\vec{x}-\vec{x}_k)^{n_k}h_k(\vec{x})
\eeq
$n_k$ being the multiplicity of that zero and $h_k(\vec{x})$ being regular and non-vanishing. As a consequence, if there are $n$ zeroes $\vec{x}_1, \vec{x}_2, \ldots , \vec{x}_n$ of $\phi$ on ${\bf R}^2$, it is possible to choose a gauge in which
\beq
\Theta (\vec{x}) =2\sum_{k=1}^n\arg (\vec{x}-\vec{x}_k)
\eeq
 $\arg (\vec{x})$ being the polar angle of $\vec{x}$. Then, for $|\vec{x}|\rightarrow\infty , \Theta (\vec{x})\simeq 2n\arg \vec{x}$ and we conclude that the vorticity of a self-dual configuration coincides with the number of zeroes of $\phi$ on ${\bf R}^2$. If this configuration is a non-topological one that has the asymptotic behaviour $|\phi (\vec{x})|\sim|\vec{x}|^{-\alpha}$, i.e. it has a zero of multiplicity $\alpha$ on $\partial{\bf R}^2$, for large $|\vec{x}|$ we can write
\beq
\phi (\vec{x})\simeq K|\vec{x}|^{-\alpha}e^{in\arg \vec{x}}
\eeq
Plugging this into (\ref{eq: 9}) one checks that the vector field has the following asymptotic polar expression
\beq
eA_r(\infty,\theta)=0, \; \; \;  eA_\theta (\infty,\theta)=-( \alpha +n)
\eeq
and hence,
\beq
\frac{e\Phi_M}{2\pi}={\rm number\,  of\;  zeroes\;  of\;  }\phi\;  {\rm on\;  }\; {\bf R}^2\cup\{\partial{\bf R}^2\}\label{eq: n87}
\eeq

Here, we do not repeat the exact computation of the index of the deformation operator of self-dual solutions of the CSH system on ${\bf R}^2$. This has been done by the authors in Reference \cite{bib:wi1} for the richer case of a doublet of Higgs fields, allowing for semi-local topological defects. The differences with respect to the cylinder case are due to the lower bound in the magnetic field of self-dual solutions in ${\cal C}_0 (v)$ and the non-existence of ``kink" solutions. Using (\ref{eq: n87}), dim${\cal M}$ can be restated as follows:
\beq
\dim{\cal M}=2[\frac{e\Phi_M}{2\pi}]-2=2[{\rm number\,  of\;  zeroes\;  of\;  }\phi\;  {\rm on\;  }\; {\bf R}^2\cup\{\partial{\bf R}^2\}]-2\label{eq: 18}
\eeq
A physical picture of the local structure of the non-topological vortex moduli space on ${\bf R}^2 $ arises: the dimension of ${\cal M}$ is associated with the number of translational degrees of freedom of the zeroes of $\phi $. In this picture, some of the asymptotic zeroes are free to move into the interior of ${\bf R}^2 $, with no  cost in energy. In this process non-topological vortices can split away from the non-topological defect. There are two restrictions to this migration of the asymptotic zeroes: 1) only zeroes of integer multiplicity are admissible on ${\bf R}^2$; 2) the lower bound on $\alpha $ shown in Section \S 2 for the existence of non-topological defects must be respected: a zero of at least quadratic order always exists on $\partial{\bf R}^2$. The natural interpretation is as follows: a non-topological defect is a composite object formed by several topological vortices and a basic non-topological lump or soliton that has two translational degrees of freedom. These fundamental non-topological lumps form a family parametrized by a constant $\omega\in (0, 1)$, carrying a magnetic flux $(2+\omega)\frac{e}{2\pi}$ and decreasing at infinity as $r^{-(2+\omega)}$. The outcome of the index theorem can be interpreted as follows:
\begin{itemize}
\item If the magnetic flux of the radial non-topological soliton is such that ${\displaystyle 2<\frac{e\Phi_M}{2\pi}<3}$, it corresponds to a indivisible non-topological lump, with an arbitrary center and two degrees of freedom.
\item When the magnetic flux is large enough, ${\displaystyle \frac{e\Phi_M}{2\pi}>3}$, the soliton can split in ${\displaystyle [\frac{e\Phi_M}{2\pi}]-2}$ vortices located away from the origin and a non topological lump of type ${\displaystyle <\frac{e\Phi_M}{2\pi}>}$; the number of translational modes is ${\displaystyle 2([\frac{e\Phi_M}{2\pi}]-2)+2=2[\frac{e\Phi_M}{2\pi}]-2}$, in consonance with the index theorem and the results of \cite{bib: jlw} for $\dim{\cal M}$. 

\end{itemize}
A rigorous proof of this decription would require to show that the vortex equation in ${\bf R}^2$
\beq
\Delta u+m^2e^u[1-e^u]=4\pi\sum_{k=1}^{[\alpha]-2}\delta(\vec{x}-\vec{x}_k)
\eeq
with the boundary condition
\beq
\frac{u(\vec{x})}{2}\simeq-(2+<\alpha>)\ln|\vec{x}|\ \ \ {\rm for}\ \ \ |\vec{x}|\rightarrow\infty
\eeq
admits a unique solution for any choice of $\alpha>2$ and $\vec{x}_k\in{\bf R}^2$. To achieve this goal is a piece of Partial Differential Equations theory out of the main trend of this paper. Instead of trying to solve this PDE problem, we now study the spectrum of the deformation operator of a non-topological vortex on a sphere. The reason for doing so is that one can perturb the equations around 
a solution with zeroes of the Higgs field located at one of the poles of the sphere in order to understand the geometrical origin of the zero modes.
We briefly recall some facts concerning abelian gauge field theory on the sphere, a formalism that we can use to properly transform the fields of the soliton in order to place the asymptotic zero at the origin of the coordinate system. Let us consider a $S^2$ covered by the two charts $U_N=\{ (\Omega , \psi )/0\leq\Omega <\pi\}$ and $U_S=\{ (\Omega , \psi )/0<\Omega\leq\pi\}$. Each chart is mapped to ${\bf R}^2$ by means of the stereographic projections:
\beq
\begin{array}{cccc}f_N: &U_N&\rightarrow&\Pi_N\simeq{\bf R}^2\\ &(\Omega , \psi )&\rightarrow&(r, \theta )=(\tan\frac{\Omega}{2}, \psi )\end{array}\; \; \;  \begin{array}{cccc}f_S: &U_S&\rightarrow&\Pi_S\simeq{\bf R}^2\\ &(\Omega , \psi )&\rightarrow&(\rho , \chi )=(\cot\frac{\Omega}{2}, -\psi )\end{array}  ,
\eeq
Excluding the origins in $\Pi_N$ and $\Pi_S$, there is a bijective homeomorphism
\beq
\begin{array}{cccc}f=f_S\circ i \circ f_N^{-1}: &\Pi_N&\rightarrow&\Pi_S\\ &(r, \theta )&\rightarrow&(\rho , \chi )=({\displaystyle \frac{1}{r}}, -\theta )\end{array}
\eeq
where $i:U \rightarrow U$ is the identity map in $ U_N\cap U_S$.

Let $L(S^2,{\bf C})$ be a line bundle over $S^2$ with fibre ${\bf C}$; the Higgs field is a section in the complex line bundle $L$ and the gauge potential defines  a connection by choosing the horizontal sub-spaces orthogonal to each fibre ${\bf C}$ in $L$. Field configurations are given locally by functions and vector fields in each trivialization, $(A_N, \phi_N)\in U_N \times {\bf C}$ and $(A_S, \phi_S)\in U_S \times {\bf C}$
and globally fixed by gluing them onto $U_N\cap U_S$ through a gauge transformation:
\begin{eqnarray}
\phi_S(\Omega , \psi )&=&e^{-in\psi}\phi_N(\Omega , \psi )\nonumber\\A_S(\Omega , \psi )&=&A_N(\Omega , \psi )+\frac{n}{e}d\psi\label{eq: 19}
\end{eqnarray}
The homotopy class of the transition function $g_{NS}^{(n)}=e^{in\psi}$ classifies the non-equivalent $L$-bundles over $S^2$ in $\Pi_1(U_N\cap U_S)=\Pi_1(S^1)={\bf Z}$ different classes.
The stereographic projections $(f_N A_N,f_N \phi_N)=(A,\phi) , (f_S A_S,\phi_S)=(C,\varphi)$ allow us to work in $\Pi_N$ and $\Pi_S$ rather than in $U_N$ and $U_S$. The field configurations $(A,\phi)$, $(C,\varphi)$ are defined in ${\bf R}^2$ and the transition functions become: 
\begin{eqnarray}
C_\rho(\rho , \chi )&=&-\frac{1}{\rho^2}A_r(\frac{1}{\rho}, -\chi )\nonumber\\C_\chi (\rho , \chi )&=&-A_\theta(\frac{1}{\rho}, -\chi ) -\frac{n}{e}\label{eq: 20}\\\varphi(\rho , \chi )&=&e^{in\chi}\phi (\frac{1}{\rho}, -\chi )\nonumber
\end{eqnarray}
Moreover, if $(A, \phi )$ satisfies the Bogomolnyi equations (\ref{eq: 9}), $(C, \varphi )$ is a solution of
\begin{eqnarray}
\frac{e}{\rho}F_{\rho\chi}&=&\frac{m^2}{2\rho^4}|\varphi|^2(|\varphi|^2-1)\nonumber\\D_\rho\varphi&+&\frac{i}{\rho}D_\chi\varphi=0   ,\label{eq: 21}
\end{eqnarray}
i.e.  a self-dual configuration on $\Pi_N$ transforms into a self-dual configuration on $\Pi_S$, corresponding to a non-Euclidean metric, the reason being that the stereographic maps used in the transformation are not isometries. 

In particular, if $(A, \phi )$ is a non-topological soliton without vorticity we are in the trivial bundle over $S^2$ i.e. we must select $n=0$ in (\ref{eq: 20}). The radial ansatz (\ref{eq: 9x}) applied to the fields on $\Pi_S$ reads:
\beq
\varphi (\rho , \chi )=h(\rho ), \; \; \; \;  C_\rho(\rho , \chi )=0, \; \; \; \;  C_\chi(\rho , \chi )=\frac{c(\rho )}{e}
\eeq
with
\beq
h(\rho )=g(\frac{1}{\rho}), \; \; \; \;  c(\rho )=-a(\frac{1}{\rho})
\eeq
Using (\ref{eq: 10}) and (\ref{eq: 10x}) we find:
\begin{eqnarray*}
h(\rho )\simeq\rho^\alpha\; \; \; \; &c(\rho )\simeq\alpha\; \; \; \; &{\rm for}\; \rho\simeq 0\\h(\rho )\rightarrow e^{\frac{u_0}{2}}\; \; \; \; &c(\rho )\rightarrow 0\; \; \; \; &{\rm for}\;  \rho\rightarrow\infty    ;
\end{eqnarray*}
the zero of multiplicity $\alpha$ initially located at the boundary of $\Pi_N$ now sits at the origin of $\Pi_S$.

From the perspective of $\Pi_S$, study of the self-dual deformations of the non-topological soliton is tantamount to consideration of the deformations of $(C, \varphi )$ preserving (\ref{eq: 21}). We start by introducing ${\displaystyle \varphi =ve^{\frac{1}{2}(w+i\vartheta )}}$ and write (\ref{eq: 21}) out of the origin as
\begin{eqnarray*}
e\nabla^2w&=&\frac{m^2}{\rho^4}e^w(e^w-1)\\eC_\rho&=&-\frac{1}{2}(\partial_\rho\vartheta +\frac{1}{\rho}\partial_\chi w)\\eC_\chi&=&-\frac{1}{2}(\partial_\chi\vartheta -\rho\partial_\rho w)   .
\end{eqnarray*}
Now, we come to the analysis of the linear perturbations of the defect following standard techniques \cite{bib: rub}. A small self-dual deformation $(\varphi^\prime , C^\prime )=(\varphi +\delta\varphi , C+\delta C)$ such that
\begin{eqnarray}
\varphi&=&\varphi +\varphi\eta\nonumber\\eC_\rho^\prime&=&eC_\rho-\frac{1}{2}(\partial_\rho\delta\vartheta +\frac{1}{\rho}\partial_\chi \delta w)\label{eq: 21x}\\eC_\chi^\prime&=&eC_\chi -\frac{1}{2}(\partial_\chi\delta\vartheta -\rho\partial_\rho \delta w)\nonumber
\end{eqnarray}
with $\eta =\frac{1}{2}(\delta w+i\delta\vartheta )$, satisfies (\ref{eq: 21}) to the first order in the perturbation if $\eta$ is a solution of the linearized equations. 
\beq
e\nabla^2\eta =\frac{m^2}{\rho^4}e^w(2e^w-1)\eta\label{eq: 21xx}
\eeq
where, to avoid gauge modes we have chosen the following background gauge condition
\beq
e\partial_kC_k^\prime =\frac{m^2}{2\rho^4}\varphi^{\prime 2}(\varphi^{\prime 2}-1)(\vartheta^\prime -\vartheta )
\eeq
Expanding $\eta (\rho , \chi )$ in Fourier series
\beq
\eta (\rho , \chi )=\sum_{k=-\infty}^{+\infty}\eta_k(\rho )e^{-ik\chi}
\eeq
we see from (\ref{eq: 21xx}) that the Fourier coefficients must solve the Schr\"{o}dinger equation
\beq
\frac{d^2\eta_k}{d\rho^2}+\frac{1}{\rho}\frac{d\eta_k}{d\rho}+[\frac{k^2}{\rho^2}-\frac{m^2}{e\rho^4}e^w(2e^w-1)]\eta_k=0\label{eq: 22}
\eeq
Because the zero of $\varphi$ at the origin has multiplicity greater than two, the centrifugal term dominates over the potential contribution in (\ref{eq: 22}) for small $\rho$. Near the origin we find the approximate solution:
\beq
\eta_k(\rho )\simeq A_k\rho^k+B_k\rho^{-k}\label{eq: 23}
\eeq
In first place, we can generically discard the existence of a regular zero mode as $\eta_0(\rho)$. Fourier modes with $k<0$ give rise to perturbations $\delta C$ in the gauge field that are singular at the origin, as one can check by an easy computation after substitution of (\ref{eq: 23}) in (\ref{eq: 21x}). If $k>0$, however, the perturbation $\delta C$ of the gauge field has a zero when $\rho\rightarrow0$. Therefore, we reject modes with $k<0$ and realize that near $\rho=0$ perturbations with $k>0$ behave as $\delta\varphi\simeq\rho^{\alpha-k}$. Perturbations such that $k\leq[\alpha ]-2$ do not take the non-topological soliton out of its moduli space because a non-topological defect has a zero of at least multiplicity 2 at the origin of $\Pi_S$ (infinity of $\Pi_N $). There are $[\alpha]-2$ allowed values of $k$, $k=1,2,...,[\alpha]-2$, and $2[\alpha]-4$ linearly independent solutions because the relation between the two complex coefficients $A_k$ and $B_k$ can be fixed by normalizability. 

The geometric meaning of these modes is clear: from (\ref{eq: 23}) and (\ref{eq: 21x}), we read the deformation produced by $\eta_k(\rho)$ near the origin:
\beq
\varphi^\prime (\rho , \chi )\simeq\rho^\alpha (1+B_k\rho^{-k}e^{-ik\chi})=\rho^{\alpha -k}e^{-ik\chi}[\rho^ke^{ik\alpha}+B_k]  .
\eeq
The new self-dual solution $\varphi^\prime$ is zero at the $k$ points 
\bdm
\rho =\sqrt[k]{|B_k|},\;\;\;\,\;\;\;\;\chi_l=\frac{\arg B_k}{k}+2\pi\frac{l}{k},\;\;\; l=1, 2, \ldots , k
\edm
away from the origin, but a zero of order $\alpha -k$ still remains at $\rho=0$. The $k^{th}$ mode describes the symmetric ``mitosis`` of $k$ asymptotic vortices from the original vorticial non-topological defect, leaving a non-topological vortex of magnetic flux $\alpha-2$ as debris (a basic non-topological soliton  for $k=[\alpha]-2$). Besides these dispersive modes, there are another two translational modes due to the freedom of choosing any point $\Pi_N$ as the center of the initial defect. The total number of degrees of freedom is $2[\alpha ]-2$, in agreement with the result of Reference \cite{bib: jlw}. 

Although our analysis has proved the local structure of the moduli space of self-dual solutions, it can be extended to the whole sphere if the fractionary part of $\alpha$ is zero. The transition functions (\ref{eq: 19}) do this job. Therefore, topological vortex solutions exist on $S^2$ and the structure of the moduli space is inferred from the analogous space in ${\bf R}^2$. When $<\alpha>$ is other than zero, however, the best thing that we can do is to push the transition region to a small circle around the North Pole (the origin in $\Pi_N$). Thus, non-topological self-dual defects only exist on a punctured sphere, $S^2-\{\infty\}$ and cannot be extended to $S^2$. A warning: to complete the argument above we should undo the stereographic transformation to obtain the vortex equations on the Riemann sphere with the standard metric.

The same arguments apply also to the degrees of freedom of the self-dual solutions on the cylinder. The physical interpretation of the index theorem result in the sector ${\cal C}_v^v$ is clear. Applying
isometrically the cylinder to ${\bf R}^2-\{(0,0)\}$ with the metric $ds^2={\displaystyle\frac{1}{r^2}}(dr^2+r^2 d\theta^2 )$
by means of the map $f:(x_1,x_2) \rightarrow (r,\theta)= (e^{x_1},x_2)$, the self-duality equations become:
\beq
\Delta u+ \frac{m^2}{|\vec{x}|^2}{e^u}[1-{e^u}]=4\pi \sum_{k=1}^n \delta(\vec{x}-\vec{x}_k)\label{eq: 79am}
\eeq
The freedom in the choice of the positions $\vec{x}_k$ of the zeroes of $\phi$ in ${\bf R}^2-\{(0,0)\}$ is the reason
for finding dim${\cal M}=2n$ in this sector. Furthermore, the analysis carried out for the non-topological defects on the plane can be extended to describe the other sectors in ${\cal C}$ on the cylinder by only replacing ${\bf R}^2$ by ${\bf R}^2-\{(0,0)\}$. Doing so, we observe that 
\begin{displaymath}
{\rm dim}{\cal M}=2[{\rm number \ of \ zeroes \ of \ \phi\ in \ }({\bf R}^2-\{(0,0)\})\cup\partial({\bf R}^2-\{(0,0)\})]+1
\end{displaymath}
in ${\cal C}_0^v$ or ${\cal C}_v^0$. Then, the outcome of the index theorem shows the possibility of migrations of the zeroes of $\phi$ from infinity towards the interior of the cylinder and viceversa; the integer part [ ] entering the formula forbids the free motion of asymptotic zeroes of non-integer order. First, we focus on the $[{\displaystyle \frac{m}{2}}]=0$ case, in which dim${\cal M}=1$. This means that there is only a translational degree of freedom, the centre $x_1=x_{1k}$ of the axial solution. The
magnetic flux is maximun at $x_1=x_{1k}$ and decays as $e^{-<\frac{m}{2}>|x_1 |}$ when $x_1 \rightarrow \pm\infty$. The axial symmetry implies that both electric and magnetic flux are wrapped around the $x_2$ coordinate for these basic ``kink" solutions. When ${\displaystyle \frac{m}{2}}>1$ we learn from the index theorem that the axial soliton may decompose in a basic kink and ${\displaystyle [\frac{m}{2}]}$ vortices. An important difference among the ``kinks" and the vortices is that, while the former have their magnetic flux concentrated on rings associated to 1-cycles generating the homology of the cylinder, the latter give rise to flux rings that are boundaries. The $2{\displaystyle [\frac{m}{2}]}$ coordinates of the vortex centres make the count of the index theorem: dim ${\cal M}=2{\displaystyle [\frac{m}{2}]}+1$.

In ${\cal C}_0^0$ matters are similar, but now the integer part in the index theorem formula applies separately to the two borders of the cylinder, i.e. from the perspective of the annulus
\begin{eqnarray*}
{\rm dim}{\cal M}&=&2\{{\rm number \ of \ zeroes \ of \ \phi\ in \ }{\bf R}^2-\{(0,0)\}+\\&+&[{\rm number \ of \ zeroes \ of \ \phi\ in\ }\{(0,0)\}]+[{\rm number \ of \ zeroes \ of \ \phi\ in \ }\infty]\}+\\&+&1 
\end{eqnarray*}
This can be understood if one assumes that there exists a family of basic non-topological defects parametrized by a constant $\omega\in(0,<\frac{m}{2}>)$ such that the scalar field falls as $e^{\mp\omega x_1}$ when $x_1\rightarrow \pm\infty$ and ${\displaystyle \frac{e\omega}{\pi}}$ units of magnetic flux are wrapped around $x_2$. Then, for $m$ big enough, the index theorem tells us that the axial solutions with $\alpha >1 $ in ${\cal C}_0^0$ decompose in a non-topological basic defect and $2[\alpha ]$ topological vortices in neutral equilibrium with the parent  non-topological vortex, to explain the $4[\alpha ]+1$ degrees of freedom.

A rigorous proof of the previous description of the physical meaning of the coordinates of the moduli space requires to show the existence of solutions of the equation (\ref{eq: 79am}) with arbitrary $\vec{x}_k\in {\bf R}^2-\{(0,0)\}$ and
\begin{itemize}
\item $n={\displaystyle[\frac{m}{2}]}$ and boundary conditions
\begin{eqnarray*}
\frac{u(\vec{x})}{2}&\simeq&{\displaystyle<\frac{m}{2}>}\ln|\vec{x}|\ \ \ {\rm for}\ \ \ |\vec{x}|\rightarrow 0\\
\frac{u(\vec{x})}{2}&\rightarrow& 0\ \ \ {\rm for}\ \ \ |\vec{x}|\rightarrow\infty
\end{eqnarray*}
for the sector ${\cal C}_0^v$ (and similarly for the ${\cal C}_v^0$).
\item $n=2[\alpha]$ and asymptotics
\begin{eqnarray*}
\frac{u(\vec{x})}{2}&\simeq&<\alpha>\ln|\vec{x}|\ \ \ {\rm for}\ \ \ |\vec{x}|\rightarrow 0\\
\frac{u(\vec{x})}{2}&\simeq&-<\alpha>\ln|\vec{x}|\ \ \ {\rm for}\ \ \ |\vec{x}|\rightarrow\infty
\end{eqnarray*}
in ${\cal C}_0^0$. 
\end{itemize}
If solutions with this asymptotic behaviour exist, the non-topological vortices of \cite{bib: jlw} can be extended to a cylinder.

Finally, we briefly comment on the question of stability. Self-dual topological vortices on the plane or the cylinder are stable because the existence of the topological conservation law $Q_2^T={\rm const}$ in ${\cal C}_v(n)$ and ${\cal C}_v^v(n)$. The same is true in the hybrid ${\cal C}_0^v$ and ${\cal C}_v^0$ sectors of the cylinder, this time the topological superselection rule comig from $Q_1^T$. In the ${\cal C}_0$ or ${\cal C}_0^0$ sectors, however, there are not topological obstructions that prevent the decay of the defects to the vacuum. Nevertheless, we know that the $U(1)$ symmetry is unbroken in the symmetric phase and the electric charge is conserved. The relation between energy and charge is exactly the same for the self-dual solitons and the fundamental quanta:
\beq
Q^{NTP}=-\frac{2\kappa}{ev^2}{\rm sign}(\Phi_M)E^{NTP} \ \ \ \ \ \ Q_{\phi}=\mp\frac{2\kappa}{ev^2}m_\phi
\eeq
so that the non-topological defects are at the threshold of instability: the solitons are in neutral equilibrium with $Q^{NTP}$ quanta (note that we are working at the classical level and therefore the ``number" of particles is not quantized).
\section{Conclusions. }
In this paper we have found examples of exact non-topological solitons in the CSH model and have attempted to clarify a mathematical conundrum (the pathology of the index theorem) and a physical one (the interpretation of non-topological objects). A picture of the non-topological sectors as containing only one non-topological lump and a number of vortices has been elaborated. As a final remark we mention that this concept reduces the question of studying the low-energy scattering of self-dual defects to investigating the vortex-vortex and vortex-lump interactions when the fields are allowed to vary slowly in time. A complete image of the dynamics in the full CSH model could be given by a suitable generalization of the methods developed in \cite{bib: sct} for the topological sector. Note, however, that as demonstrated in these papers the usual adiabatic approximation is not satisfactory for the CSH system. An imaginative work of Manton \cite{bib: fovd} has stimulated the search for models that, giving rise to the same static solutions, have more tractable dynamics \cite{bib: ult}. This type of models could also give valuable insight into the subject of vortex-lumps forces, perhaps by using methods similar to those of \cite{bib: jcr}.

\newpage
\begin{figure}[h]
\begin{center}
\epsfig{file=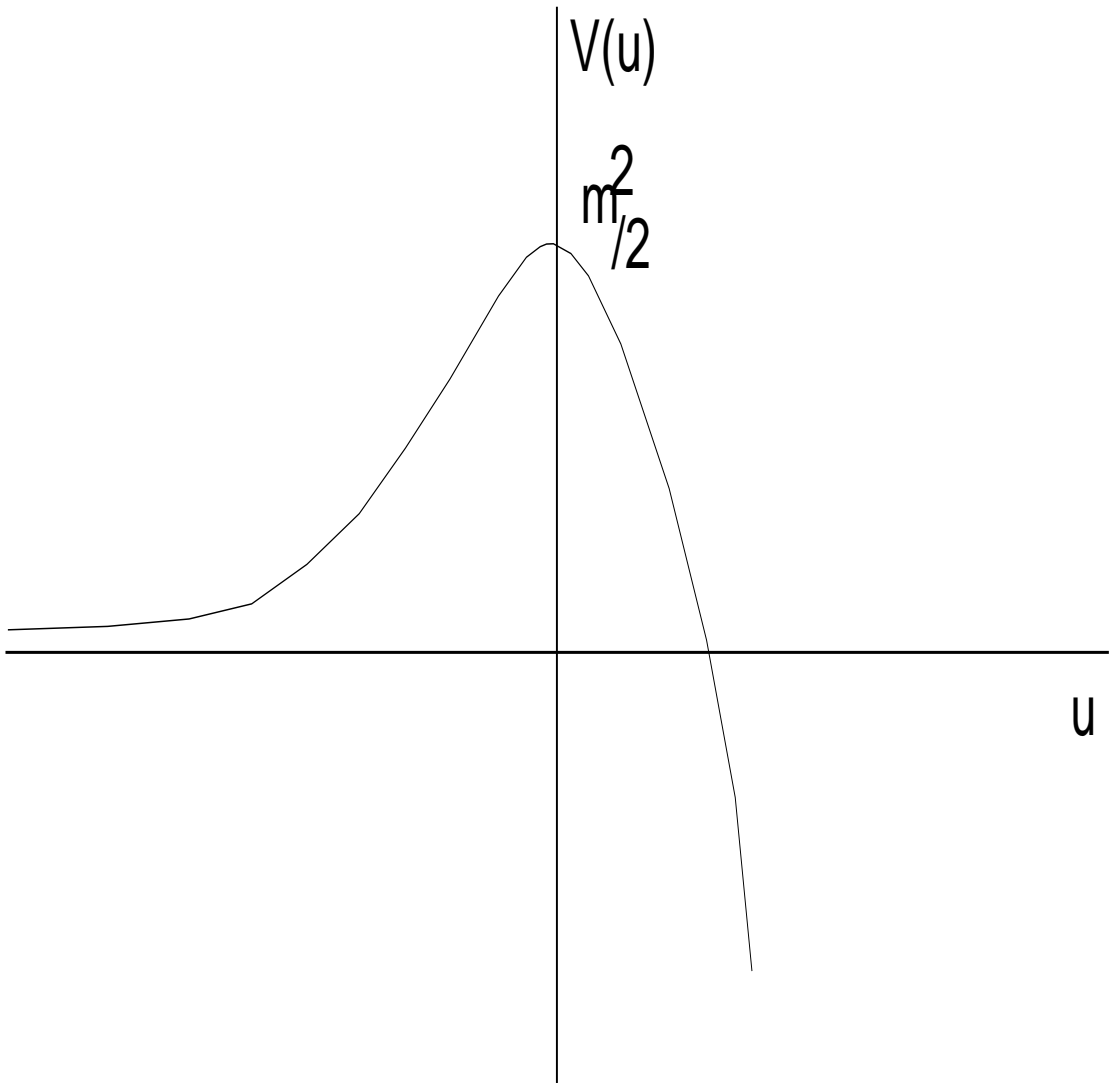,height=15cm,width=15cm}
\end{center}
\caption{The potential energy $V(u)$ for the mechanical analogy.}
\end{figure}
\newpage
\begin{figure}[h]
\begin{center}
\epsfig{file=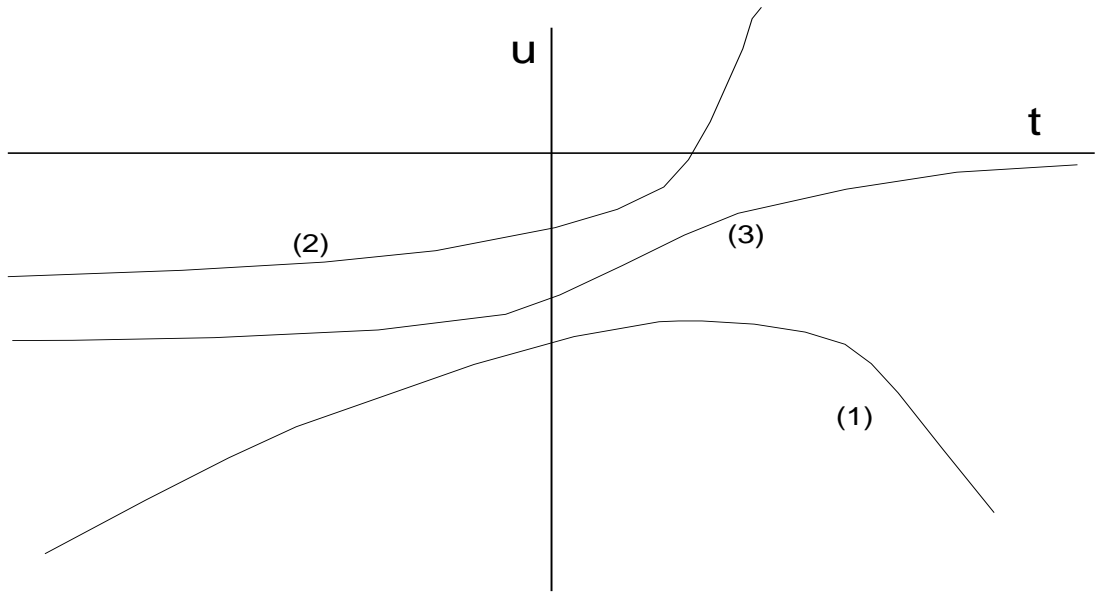,height=8cm,width=15cm}
\end{center}
\caption{The evolution of the mechanical analogs of the radial defects on the plane.}
\end{figure}
\newpage
\begin{figure}[h]
\begin{center}
\epsfig{file=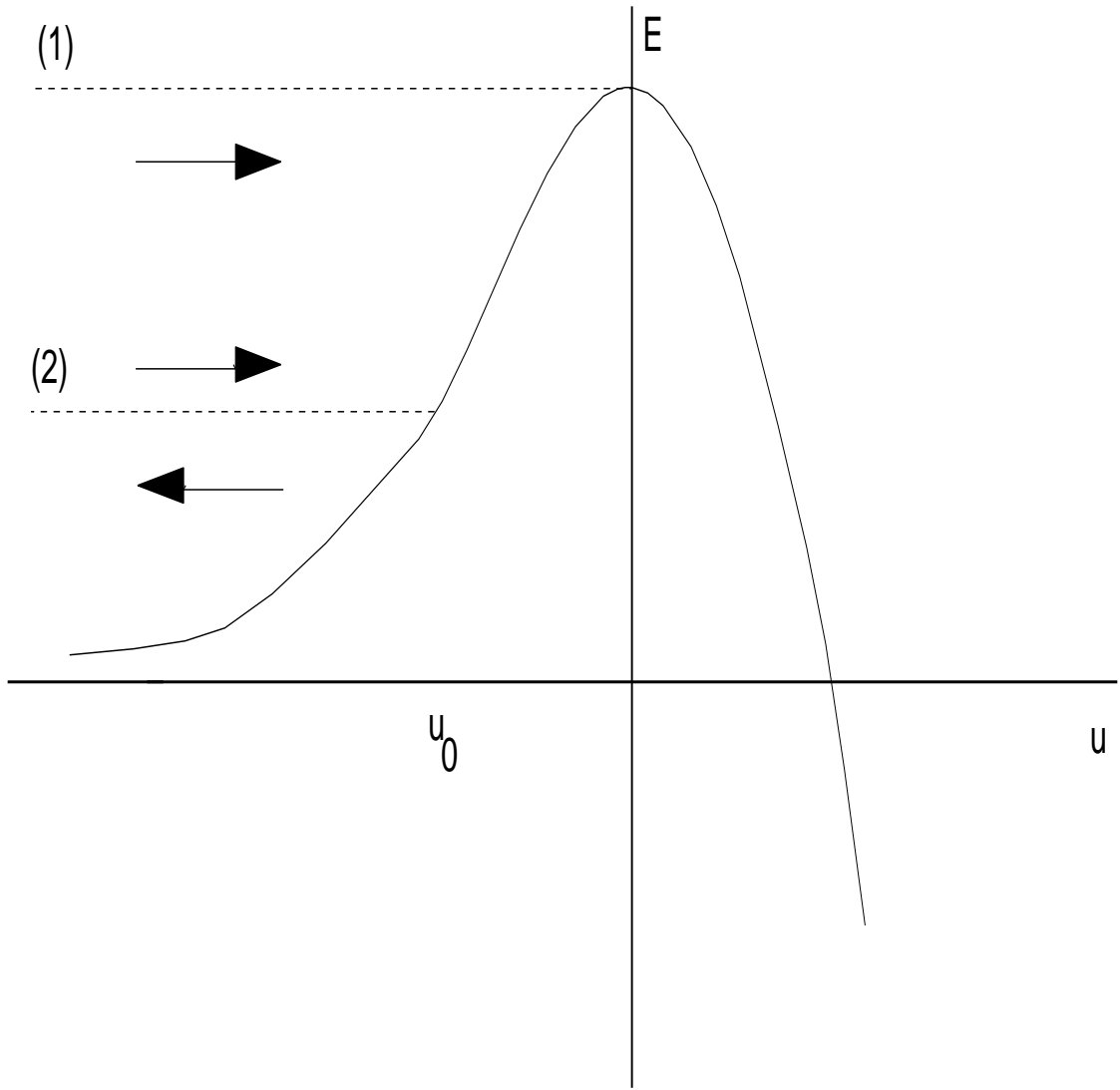,height=15cm,width=15cm}
\end{center}
\caption{The $u-{\cal E}$ plot for the mechanical analogs of the axial defects on the cylinder. (1): solution in ${\cal C}_0^v$; (2): solution in ${\cal C}_0^0$}
\end{figure}
\end{document}